\documentclass[usenatbib, letters]{mnras}

\usepackage[T1]{fontenc}
\usepackage{ae,aecompl}

\usepackage{graphicx}	

\title[TXS 1206+549: a $\gamma$-ray detected NLS1]{TXS 1206+549: a new $\gamma$-ray detected narrow-line Seyfert 1 galaxy at redshift 1.34?}

\author[Rakshit et al.]{Suvendu Rakshit$^{1,2}$\thanks{E-mail: suvenduat@gmail.com},
Malte Schramm$^{3}$, C. S. Stalin$^{4}$, I. Tanaka$^{5}$, 
Vaidehi S. Paliya$^{1,6}$,
\newauthor
Indrani Pal$^{4}$, 
Jari Kotilainen$^{2,7}$, Jaejin Shin$^{8}$
\\\\
$^{1}$ Aryabhatta Research Institute of Observational Sciences, Manora Peak, Nainital 263002, India \\
$^{2}$ Finnish Centre for Astronomy with ESO (FINCA), University of Turku, Quantum, Vesilinnantie 5, 20014, Finland \\
$^{3}$ Graduate school of Science and Engineering, Saitama Univ. 255 Shimo-Okubo, Sakura-ku, Saitama City, Saitama 338-8570, JAPAN \\
$^{4}$ Indian Institute of Astrophysics, Block II, Koramangala, Bangalore-560034, India \\
$^{5}$ Subaru Telescope, National Astronomical Observatory of Japan, 650 North A'ohoku Place, Hilo, Hawaii, 96720, U.S.A. \\
$^{6}$ Deutsches Elektronen Synchrotron DESY, Platanenallee 6, 15738 Zeuthen, Germany \\
$^{7}$ Tuorla Observatory, Department of Physics and Astronomy, FI-20014 University of Turku, Finland \\
$^{8}$ Department of Astronomy and Atmospheric Sciences, Kyungpook National University, Daegu 41566, Republic of Korea \\
}

\begin{document}

\pagerange{\pageref{firstpage}--\pageref{lastpage}} \pubyear{2020}

\maketitle

\begin{abstract}
Radio and $\gamma$-ray loud narrow-line Seyfert 1 galaxies (NLS1s)  are 
unique objects to study the formation and evolution of relativistic jets, as 
they are believed to have high accretion rates and powered by low mass black 
holes contrary to that known for blazars. However, only about a dozen 
$\gamma$-ray detected NLS1s  ($\gamma$-NLS1s) are known to date and all of them are at $z\le1$. 
Here, we report the identification of a new $\gamma$-ray emitting NLS1 TXS 1206+549 at $z=1.344$. A near-infrared spectrum taken with the Subaru telescope showed H$\beta$ emission line with FWHM of $1194\pm77$ km s$^{-1}$ and weak [O III] 
emission line but no optical Fe II lines, due to the limited wavelength coverage and poor signal-to-noise ratio. However, UV Fe II lines are present in the SDSS optical spectrum. The source is very radio-loud, unresolved, and has a flat radio spectrum. The broadband SED of the source has the typical two hump structure shown by blazars and other $\gamma$-NLS1s. The source exhibits strong variability at all wavelengths such as the optical, infrared, and $\gamma$-ray bands. All these observed characteristics show that TXS 1206+549 is the most distant $\gamma$-NLS1 known to date.     
\end{abstract}


\section{Introduction}

The physical processes required to launch powerful relativistic jets 
in active galactic nuclei (AGN) are still unclear inspite of observational advances across wavelengths in the recent years. According to \citet{2000ApJ...543L.111L}, AGN hosted in elliptical galaxies with black hole (BH) masses $>$ $10^8 M_{\odot}$ can produce large scale relativistic jets
while AGN hosted in late-type galaxies with less massive  
BHs are mostly unable to produce relativistic jets. 
This suggests that massive BHs are needed to launch powerful relativistic 
jets. However, the detection of $\gamma$-ray emission 
\citep{2009ApJ...707L.142A,2018ApJ...853L...2P} from a few radio-loud narrow-line Seyfert 1 galaxies (NLS1s) unambiguously argues for the 
presence of closely aligned relativistic jets in them similar to 
blazars \citep{2008ApJ...685..801Y}. This therefore, challenges our understanding of how relativistic 
jets are formed, as NLS1s are usually powered by low mass BHs in 
late-type galaxies \citep{2020MNRAS.492.1450O}, while blazars are believed to be powered by high mass 
BHs in elliptical galaxies.

 \begin{figure*}
  \resizebox{17.5cm}{7cm}{\includegraphics{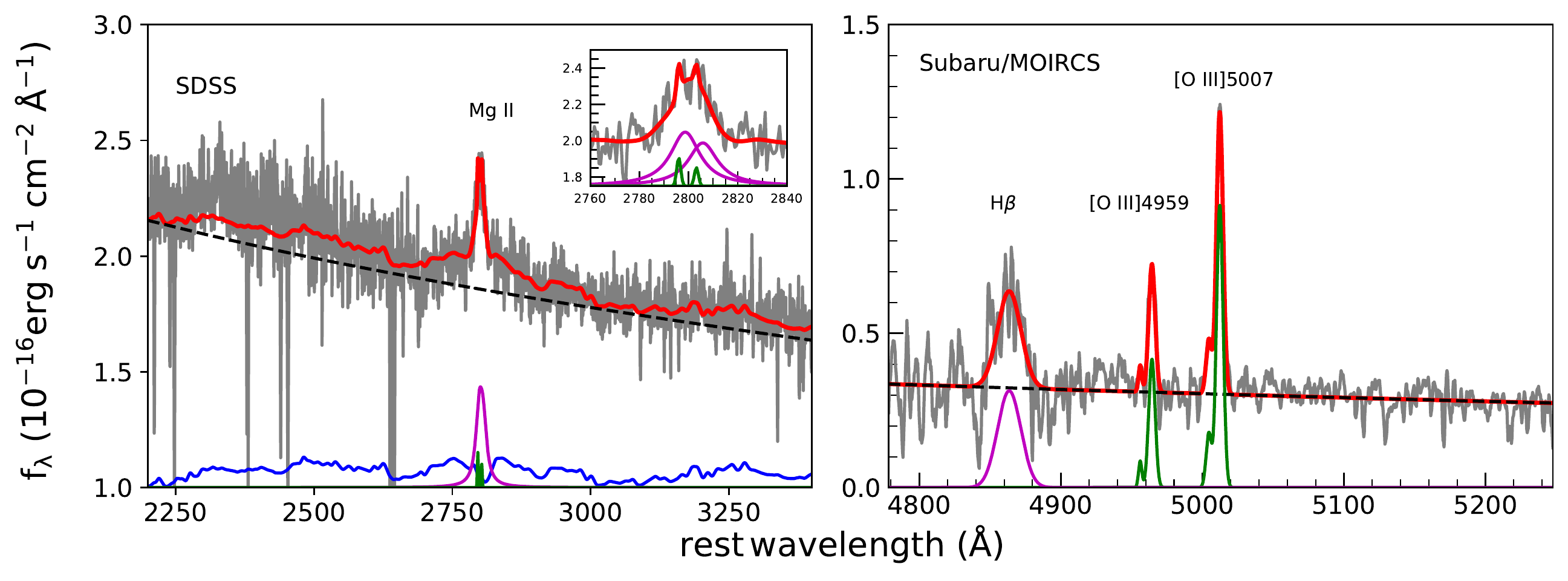}}  
\caption{Spectrum of TXS 1206+549. Left: Mg II line fit of the SDSS spectrum for MJD=57430. The data (gray), best-fit model (red) and power law plus Balmer continuum 
(black dashed) are shown along with the decomposed broad (magenta) and narrow (green) components of Mg II, and 
Fe II in UV (blue). The inset plot shows the zoomed version of Mg II line fitting. Right: H$\beta$ model fit to the Subaru spectrum.  The 
data (gray), best-fit model (red), power law continuum (black dashed), 
decomposed broad line (magenta) and narrow lines (green) are 
shown.}\label{Fig:spectra} 
 \end{figure*}

NLS1s \citep{1985ApJ...297..166O} constitute a unique 
class of AGN having a low BH mass accreting close to the Eddington limit \citep{2000PASJ...52..499M,2017ApJS..229...39R}, showing strong optical Fe II emission, soft X-ray excess \citep{2020ApJ...896...95O} and lower amplitude of optical variation \citep{2017ApJ...842...96R} compared to their broad line counterparts. Although only 5\% NLS1s are radio-loud \citep{2006AJ....132..531K}, a small number of only 16 are found to be detected in high energy $\gamma$-rays \citep[see][and references therein]{2018ApJ...853L...2P,2019ApJ...872..169P}. It has been argued that the low BH mass of NLS1s could be due to  
orientation effects caused by the flat geometry of the broad line 
region \citep[e.g.,][]{2008MNRAS.386L..15D}.

The definition of NLS1 requires the full width at half maximum (FWHM) of the H$\beta$ emission line to be lesser than 2000 km s$^{-1}$ \citep{1985ApJ...297..166O}.
Most of the NLS1s that we know as of today 
are from large optical spectroscopic surveys and are therefore at $z<0.8$ \citep{2002AJ....124.3042W,2006ApJS..166..128Z,2017ApJS..229...39R,2018A&A...615A.167C}. At high-redshift, it is expected to detect luminous objects and hence identification of NLS1 at $z>1$ could be crucial to verify the claims that NLS1s are low-luminosity AGN as well as to understand the nature of these sources, and how they differ from the classical
Seyfert 1 galaxies. Of the NLS1s, only 16 are detected in the $\gamma$-ray 
band and among them only two are  at redshift beyond 0.8;  SDSS J1222+0413 at 
$z=0.966$ \citep{2015MNRAS.454L..16Y} and SDSS J094635.06+101706.1 at 
$z=1.004$ \citep{2019MNRAS.487L..40Y}.  Here, we report the identification of 
a $\gamma$-ray detected NLS1 TXS 1206+549 (SDSS J120854.24+544158.1) at 
$z=1.344$. This is based on the infrared spectroscopic observations carried out using the 8.2m Subaru telescope which showed the presence of H$\beta$ line with FWHM of $1194\pm77$ km s$^{-1}$ and [O III] to H$\beta$ flux ratio $\sim0.7$. The optical Fe II lines are not seen in the Subaru spectrum, however, UV Fe II lines are present in the optical spectrum from the Sloan Digital Sky Survey (SDSS) (see section \ref{sec:spectroscopy}). The 
multi-band properties of this source are presented in section 
\ref{sec:multi-band} with a discussion and conclusion in 
section \ref{sec:conclusion}. Throughout the paper, a cosmology with 
$H_0 = 70$ km s$^{-1}$ Mpc$^{-1}$, $\Omega_{\Lambda}$ = 0.7 and 
$\Omega_{M}$ = 0.3 is assumed.

  \begin{table*}
  \caption{Spectral properties of TXS 1206+549. Line widths are not corrected for instrumental resolution. The luminosity of UV Fe II in the wavelength range of $2200-3090$\AA~ is also given. Due to poor S/N, [O II]$\lambda3727$ emission line parameters for spectrum at MJD=52672 are not calculated.}
  \begin{center}
  \hspace*{-0.5cm}
  \scalebox{0.72}{
  \begin{tabular}{ccccccccc} \\ \hline \hline  
    &   &    &   & SDSS spectra &   &  &  & \\
  SDSS ID   &  model (Mg II, br)  &   FWHM (Mg II, br)  &   EW (Mg II, br)  &  flux (Mg II, br)  & flux (Fe II, UV) & $L_{3000}$ &  FWHM ([O II]$\lambda3727$) & flux ([O II]$\lambda3727$)  \\ 
       &    &   (km s$^{-1}$) & (\AA) & (erg s$^{-1}$ cm$^{-2}$) & (erg s$^{-1}$ cm$^{-2}$) & (erg s$^{-1}$) & (km s$^{-1}$) & (erg s$^{-1}$ cm$^{-2}$) \\
  \hline
  1018-52672-0120 & Gaussian   & $2975\pm 261$  & $18.7\pm  1.1$  & $133.59\pm8.11$  & $604.33\pm59.25$  & $45.295\pm 0.004$  & --  & -- \\
  				  & Lorentzian & $2560\pm 270$  & $22.1\pm  1.9$  & $158.64\pm14.05$ & $501.99\pm 66.40$  & -- & -- & -- \\
  6688-56412-0776 & Gaussian   & $2838\pm 220$  & $12.0\pm  0.9$ & $100.99\pm 7.79$  & $652.82\pm 45.19$ & $45.370\pm 0.002$ & $662\pm 158$  & $23.29\pm 4.14$ \\
                  & Lorentzian & $2353\pm 161$   & $14.5\pm  0.8$  & $122.33\pm 6.70$  & $593.31\pm 40.70$ & -- & --  & -- \\
  8229-57430-0099 & Gaussian   & $2838\pm 370$   & $5.2\pm  0.6$ & $96.62\pm 12.17$ & $666.09\pm 56.32$ & $45.718\pm 0.001$  & $584 \pm 71$ & $27.18\pm 2.82$  \\
                  & Lorentzian & $2076\pm 280$   & $6.3\pm  0.6$  & $117.22\pm 12.16$  & $619.98\pm 54.49$ & -- & -- & --  \\ \\ \hline \hline
                  
   &   &    &   & Subaru spectrum &   &  &  & \\
    &  & $\log L_{5100}$ &  Model (H$\beta$) & FWHM$(\mathrm{H\beta})$ & flux ($\mathrm{H\beta}$) & flux ([O III]$\lambda$5007)  & & \\
   &   &   (erg s$^{-1}$) &    & (km s$^{-1}$) & (erg s$^{-1}$ cm$^{-2}$) & (erg s$^{-1}$ cm$^{-2}$) \\ \hline
   &   & $45.20 \pm 0.05$  & Single Gaussian & $1194\pm77$ & $76.65\pm1.03$  & $51.96\pm0.55$\\            \hline \hline
  \end{tabular}}
  \label{Table:data}
  \end{center}
    \end{table*}

\section{Spectroscopy}\label{sec:spectroscopy} 
\subsection{Optical}
To find high-$z$ $\gamma$-ray emitting NLS1s, we cross-matched the fourth catalog of 
AGN detected by the Fermi/LAT \citep[4LAC;][]{2020_Ajello} with the spectral catalog of SDSS DR14 quasars \citep{2020ApJS..249...17R}, that provides spectral properties 
of around 500,000 quasars. We looked for potential narrow Mg II line objects with FWHM $<$ 2500 km s$^{-1}$ and showing traces of strong Fe II complex and identified a NLS1 candidate, TXS 1206+549, at $z=1.344$ with FWHM (Mg II) $=2419\pm581$ km s$^{-1}$. The redshifted H$\beta$ line falls outside of the SDSS spectral coverage, and thereby preventing a secure NLS1 classification of the source. 

Looking into the SDSS archive, we found three optical spectra of TXS 1206+549. Since Mg II doublets are clearly resolved, we carefully modeled all the three spectra. First, we fit the 2200$-$3500\AA\,AGN continuum using a combination of power-law, a Balmer pseudo-
continuum \citep{1982ApJ...255...25G} and the Fe II template from \citet{2006ApJ...650...57T} as described in \citet{2019ApJ...874...22S}. After fitting and subtracting the continuum, each of the Mg II $\lambda\lambda$2796, 2803 doublet lines were fitted with a broad and narrow components following \citet{2009ApJ...707.1334W}. The broad components were modeled with either a Gaussian or a Lorentzian. The width and velocity of the doublets were tied and the flux ratio was varied between 2:1 to 1:1. The narrow components were modeled using single Gaussian with same constrains as above. The decomposition of the SDSS spectrum for the epoch MJD=57430 is shown in the left 
panel of Figure \ref{Fig:spectra}. The best-fit spectral properties are summarized in Table \ref{Table:data}. The FWHM (Mg II) is found to be between 2000 to 3000 km s$^{-1}$ depending on the model used to represent the broad Mg II doublets and the epoch used. Broad line profile was similarly well fitted with either a Gaussian or a Lorentzian profile, however, in the latter case, a lower FWHM of Mg II lines was found hinting the NLS1 nature of TXS 1206+549. To confirm this, we carried out Subaru near-infrared (IR) spectroscopy, to detect and characterize the redshifted H$\beta$ line.

\subsection{NIR}

Subaru/MOIRCS long-slit near-IR spectroscopic observations were obtained on 
July 20, 2020, as  part of performance verification observations for the new LightSmyth J-band Grism (Tanaka et al. in prep.). Weather condition was good, 
with $\sim 0.8^{\prime\prime}$ seeing. We chose the $0.8^{\prime\prime}$-width slit, by which we got a spectral resolution of R$\sim 2200$. The exposure time was 300 sec with 12-times multi-sampling, and a standard A-B dither (4$^{\prime\prime}$ throw length) was applied. In total three sequences were observed with a total integration time of 1800 sec. The data was reduced following standard reduction procedures for long-slit spectroscopy using 
IRAF\footnote{Image Reduction and Analysis Facility}. The spectrum was smoothed by a 5-pixel box-car to increase the S/N and then brought to the rest-frame.

The Subaru spectrum, covering the rest-frame wavelength range of $4778-5247$\AA~ is shown in the right panel of Figure \ref{Fig:spectra}. To estimate the emission line parameters, we fit the spectrum using a power-law continuum, a single Gaussian for H$\beta$ line and a double Gaussian for each of the [O III]$\lambda$5007 and [O III]$\lambda$4959 doublets. Since no narrow component is clearly visible in the H$\beta$ line, we did not attempt to decompose it. Furthermore, due to insufficient wavelength coverage below 4700 \AA, the decomposition of optical Fe II was not performed. The decomposed model components are shown in Figure \ref{Fig:spectra} and the best-fit emission line parameters are given in Table \ref{Table:data}. The FWHM of H$\beta$ is $1194\pm 77$ km s$^{-1}$ and the flux ratio of [O III]$\lambda$5007 to H$\beta$ is $\sim0.7$ satisfying the criteria of a NLS1. We note significant UV Fe II emission in the SDSS spectrum but the insufficient wavelength coverage of the near-IR spectrum prevents us to discuss its optical Fe II strength.

\section{Multi-band properties}\label{sec:multi-band} 
\subsection{Radio-band}
TXS 1206+549 is detected in many radio surveys. It appears to be a point source in the radio image of VLBA Imaging and Polarimetry Survey \citep[VIPS;][]{2007ApJ...658..203H} at 5GHz with a peak ($\theta_{\mathrm{peak}}$) and integrated ($\theta_{\mathrm{int}}$) flux densities of 231.6 mJy and 264.5 mJy ($L_{\mathrm{int}}=1.2 \times 10^{34}$ erg s$^{-1}$ Hz$^{-1}$), respectively, displaying powerful Fanaroff-Riley (FR) II type radio jets. The  concentration parameter $\theta=\sqrt{\theta_{\mathrm{int}}/\theta_{\mathrm{peak}}}$ \citep{2002AJ....124.2364I} is 1.07 close to 1.06 for the source to be ``resolved'' according to \citet{2008AJ....136..684K}, indicating compact structure. Such compact and powerful radio jets of TXS 1206+549 indicate similarities with blazars.   

The spectral index is flat with $\alpha_r=0.3$ between 1.4 to 4.85 GHz \citep[$s_{\nu}\propto\nu^{\alpha_r}$;][]{1992ApJS...79..331W} and $\alpha_r=0.23$ between 147 MHz (TGSS) and 1.4 GHz (NVSS) \citep{2018MNRAS.474.5008D}. The source is known to be highly variable in the radio, with the radio flux 
varying between 160-302 mJy at 1.4 GHz within a time scale of few years. 
\citet{2018ApJ...866..137L} modeled the 15 GHz light curve of this source 
observed by the Owens Valley Radio Observatory (OVRO) as a combination  
of multiple flares and estimated brightness temperature 
$T_{\mathrm{var}}=5\times10^{14}$ K, Doppler factor 
$\delta_{\mathrm{var}}=29.09^{+10.28}_{-10.02}$ indicating high Doppler-boosted emission. We estimated the radio loudness ($R$) of this source using $R= f_{\nu}(5 \mathrm{GHz})/f_{\nu} (\mathrm{4400 \AA})$, where $f_{\nu}(5\mathrm{GHz})=252$ mJy at VLA 5GHz \citep{2012ApJ...744..177L} and $f_{\nu} (4400 \AA)$ is taken from \citet{2020ApJS..249...17R}. The source is very radio loud with $R\sim1300$. However, the radio loudness can have a large range due to high amplitude of variation seen in the optical ($\sim$2 magnitude; see next section) and radio bands (a factor of 2).

 \begin{figure}
  \resizebox{9cm}{4cm}{\includegraphics{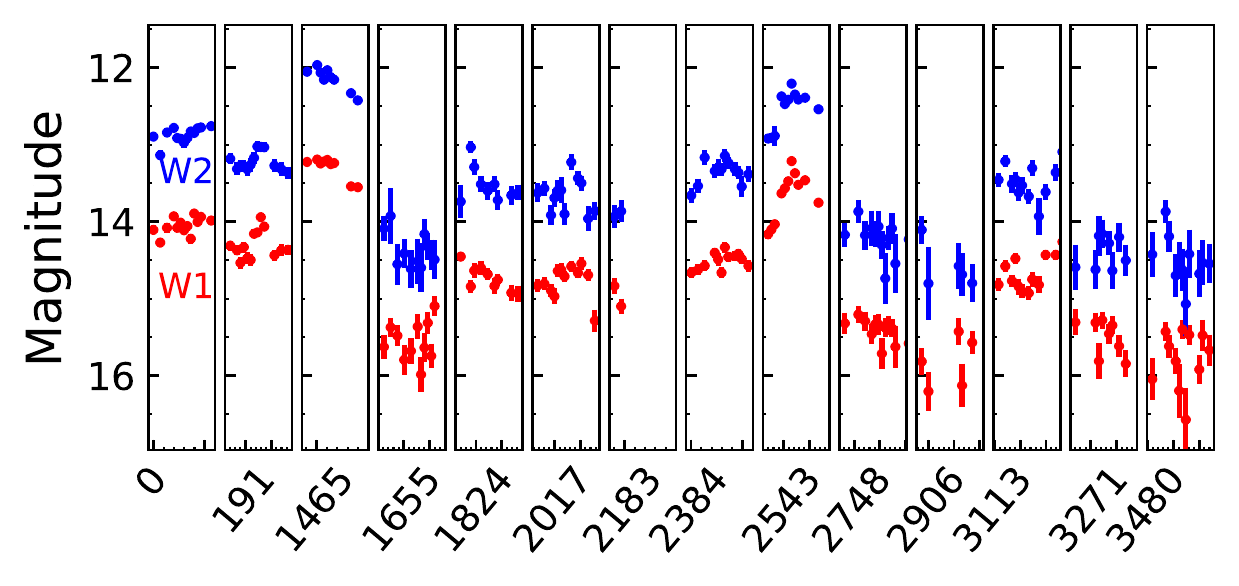}}
  \resizebox{9cm}{4.5cm}{\includegraphics{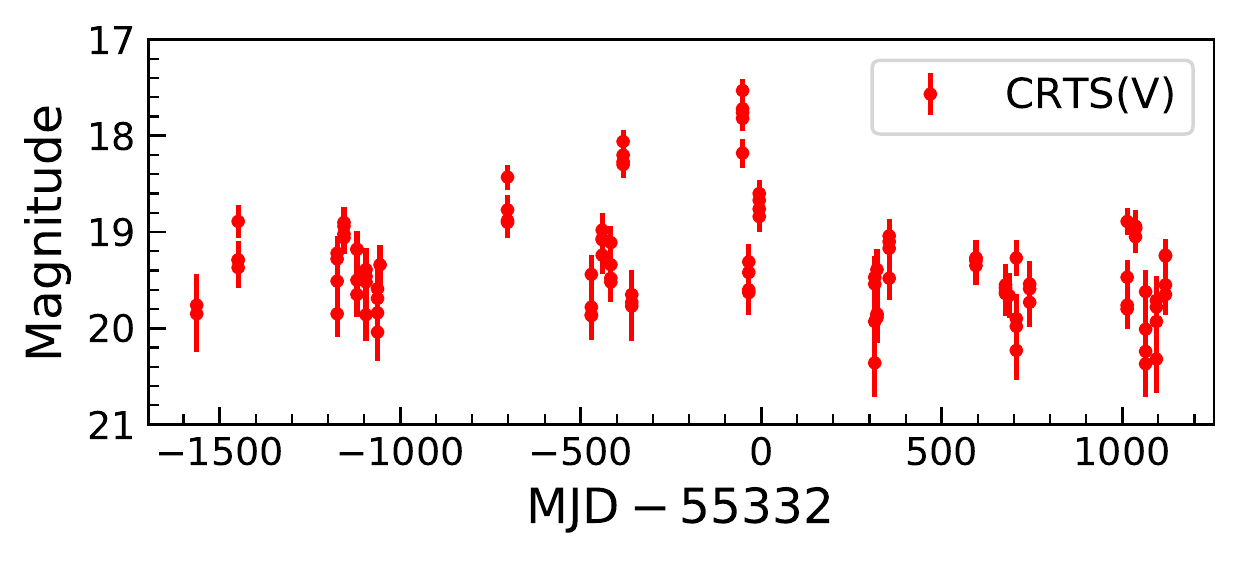}}   
  \caption{Top: WISE light curve of TXS 1206+549 in 3.4$\mu$m (W1) and 4.6$\mu$m (W2) bands with each box representing a duration of 1.2 days. Bottom: optical photometric light curve from CRTS ($V$-band).}\label{Fig:wise} 
 \end{figure}

\subsection{Infrared and optical variability}
We collected infrared photometry of TXS 1206+549 observed by the 
{\it Wide-field Infrared Survey Explorer} \citep[WISE;][]{2010AJ....140.1868W} 
from ALLWISE and NEOWISE database \citep{2014ApJ...792...30M}. The WISE 3.4$\mu$m (W1) and 4.6$\mu$m (W2) band light curves are shown in the top panel of Figure \ref{Fig:wise}. We calculated the intrinsic variability amplitude $\sigma_m$ \citep{2019MNRAS.483.2362R} i.e. the variance of the observed light curve after removing the measurement uncertainty. The duty cycle \citep[DC;][]{1999A&AS..135..477R}, which is the fraction of time when an object is variable on intra-day (i.e. 1.2 days) time-scale \citep[see also][]{2019MNRAS.483.2362R} was found to be 91\% and 79\% in W1 and W2 bands, respectively, suggesting that TXS 1206+549 is highly variable. The maximum variation was found on 2nd May 2017 with $\sigma_m=0.65\pm0.10$ mag in W1 and $0.68\pm0.11$ mag in W2 bands. Observations of such short time scale variations
in the IR indicate that the emission region is compact, much smaller than the 
dust torus, and the emission is dominated by non-thermal emission from the 
jet \citep[see][]{2012ApJ...759L..31J,2019MNRAS.483.2362R}. In the  ten years of WISE observations, TXS 1206+549 showed large amplitude variation with $\sigma_m=0.73\pm0.04$ mag in W1 and $0.69\pm0.04$ mag in W2 bands. This is similar to the variation of other $\gamma$-ray detected NLS1s 
\citep{2019MNRAS.483.2362R}.

We also looked for archival optical photometry of TXS 1206+549. The 
optical light curve using data acquired by the Catalina Real-Time Transient 
Survey \citep[CRTS;][]{2009ApJ...696..870D} is shown in the bottom panel of Figure \ref{Fig:wise}. Large amplitude flux variation with  $\sigma_m=0.52\pm0.04$ mag 
is found in the optical over the seven years of CRTS observations. This too is
consistent with the high amplitude optical variation observed in other $\gamma$-ray detected radio-loud NLS1s \citep{2017ApJ...842...96R,2019MNRAS.487L..40Y}.

\subsection{X-ray and $\gamma$-ray emission} 

TXS 1206+549 was observed by {\it SWIFT$-$XRT} \citep{2005SSRv..120..165B}
six times between March 2011 and February 2018. Among them, only
the OBSID 00040572002 has sufficient S/N for spectral analysis. For this
OBSID, we generated the spectrum using the online version of the
Swift-XRT data products
generator\footnote{\url{https://www.swift.ac.uk/user objects/}}. Fitting the spectrum
with the model TBabs*zTBabs*powerlaw we obtained a photon index of $1.7^{+0.7}_{-0.4}$, which is relatively flat compared to the radio-quiet NLS1s but similar to the $\gamma$-ray detected NLS1s \citep{2019ApJ...872..169P}. The flux in the 0.3$-$10 keV band is $7.0^{+2.8}_{-2.4} \times 10^{-13}$ ergs s$^{-1}$ cm$^{-2}$ corresponding to an apparent luminosity of $\sim 1.9 \times 10^{45}$ ergs s$^{-1}$.

TXS 1206+549 is detected at a significance level of 48$\sigma$ and associated 
with the $\gamma$-ray source 4FGL J1208.9+5441 in the 4LAC catalog 
\citep{2020_Ajello}. The $\gamma$-ray flux is $1.8\times 10^{-11}$ 
ergs s$^{-1}$ cm$^{-2}$ in the energy range of 100 MeV to 100 GeV and 
corresponds to an isotropic $\gamma$-ray luminosity of 4.8$\times10^{46}$ 
ergs s$^{-1}$. The spectrum is best-fitted by a log Parabola model with 
index $\alpha=2.51\pm0.03$ and curvature  $\beta=0.07\pm0.02$ compared to a 
power-law model ($\alpha=2.59\pm0.02$). The best-fit photon index is in the 
range of $2.2-2.8$ which is similar to that seen in other $\gamma$-ray 
detected NLS1s \citep[e.g.,][]{2009ApJ...707L.142A,2015MNRAS.454L..16Y,2019MNRAS.487L..40Y}. TXS 1206+549 is also variable in the $\gamma$-ray band with a 
fractional variability amplitude of $0.49\pm0.12$ \citep[see][]{2020_Ajello}.

 \begin{figure} 
 \resizebox{8.5cm}{8.5cm}{\includegraphics{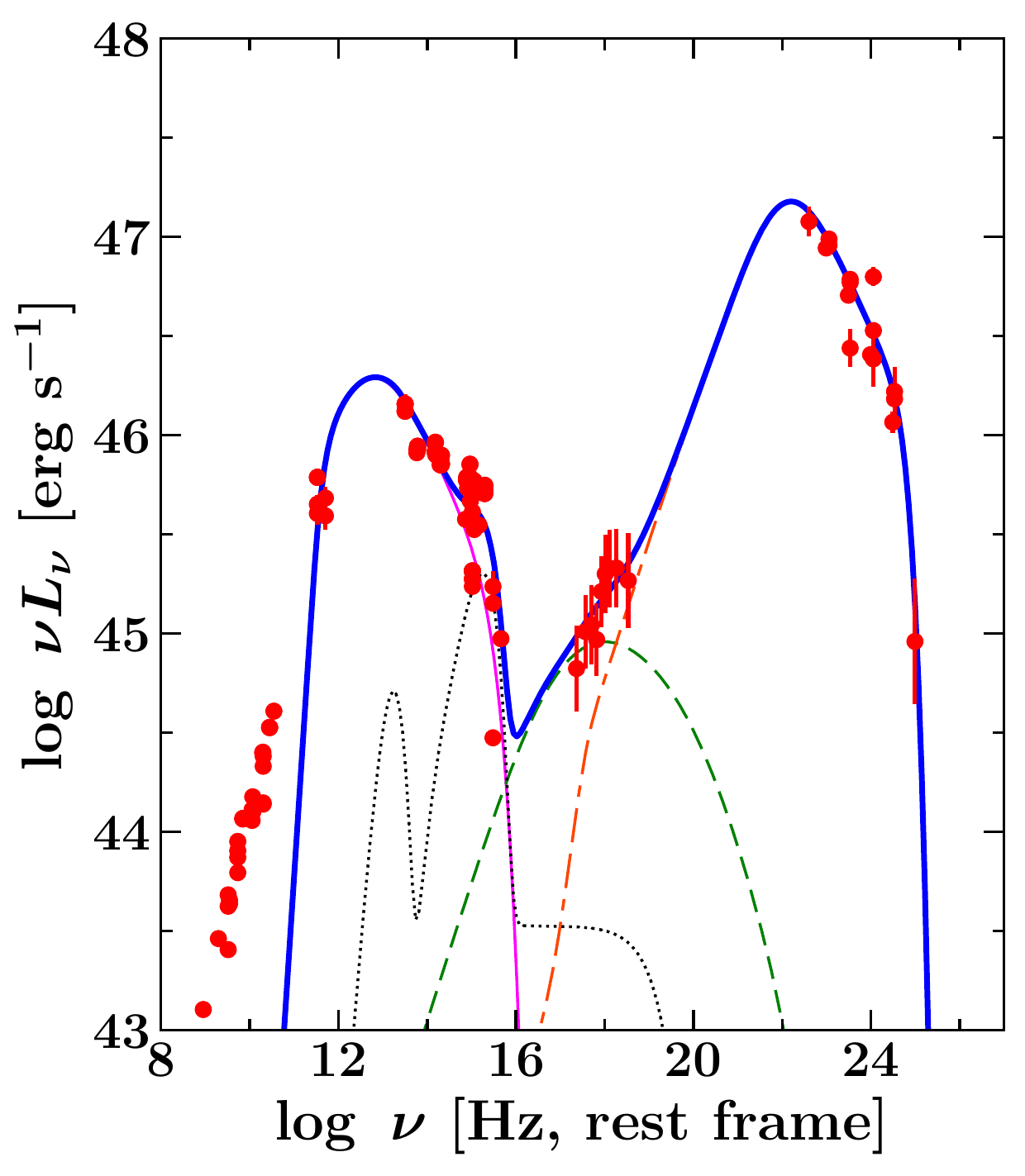}} 
  \caption{Rest frame multiwavelength SED of TXS 1206+549 along with the results of the leptonic radiative modeling. Pink thin solid, green dashed, and orange dash-dash-dot lines represent synchrotron, SSC, and EC processes, respectively. Thermal emission from the dusty torus, accretion disk, and corona is shown with the black dotted line. Blue thick solid line refers to the sum of all the radiative components.}\label{Fig:sed} 
 \end{figure}

\subsection{Broad band Spectral Energy Distribution}
We collected non-simultaneous, archival observations of TXS 1206+549 from the 
Space Science Data Center (SSDC\footnote{\url{https://tools.ssdc.asi.it/SED/}}) and 
plotted the multiwavelength spectral energy distribution (SED) in 
Figure~\ref{Fig:sed}. As can be seen, the broadband SED of TXS 1206+549 exhibits the 
typical double hump structure similar to blazars. Indeed, this object is 
classified as a flat spectrum radio quasar (FSRQ) in the 4LAC 
catalog \citep[see also][]{2020_Tan}.

To investigate the physical properties of the source, we reproduced the SED with the conventional synchrotron inverse Compton radiative model \citep[see details in][]{2009ApJ...692...32D,2009MNRAS.397..985G}. The associated results are plotted in Figure~\ref{Fig:sed} and the derived parameters are provided in Table~\ref{tab:sed}. For the modeling, we used the geometric mean of the black hole 
masses computed from the H${\beta}$ and Mg~II emission lines (see next section). The broad line region (BLR) luminosity ($L_{\rm BLR}$), on the other hand, was estimated from the line luminosities derived from the spectral fitting and assuming the line scaling factors reported in \citet[][]{1991ApJ...373..465F} and \citet[][]{1997MNRAS.286..415C}. Assuming a 10\% BLR covering factor, we calculated the accretion disk luminosity from $L_{\rm BLR}$.

The high-frequency radio to optical-UV spectrum is found to be well explained by synchrotron emission with a minor contribution from the accretion disk. The X- and $\gamma$-ray spectra are reproduced by a combination of the synchrotron self Compton (SSC) and external Compton (EC) processes. This suggests a similarity of the physical properties of TXS 1206+549 with FSRQ type blazars. Furthermore, the derived SED parameters (Table~\ref{tab:sed}) are consistent with that typically determined for broad line blazars detected with the {\it Fermi}-LAT \citep[][]{2017ApJ...851...33P}.

\begin{table}
\caption{Summary of the parameters used/derived from the modeling of the SED.}\label{tab:sed}
\begin{tabular}{lc}
\hline
\hline 
Parameter & Value \\
\hline
Slope of the broken power law before break energy & 1.6\\
Slope of the broken power law after break energy  & 4.0 \\
Minimum Lorentz factor  & 1 \\
Break Lorentz factor & 206  \\
Maximum Lorentz factor & 4200 \\
Magnetic field, in Gauss & 3.2 \\
Bulk Lorentz factor & 11\\
Compton dominance & 8 \\
Distance of the emission region, in parsec & 0.05 \\
Size of the BLR, in parsec & 0.05 \\
Log scale black hole mass, in solar mass unit &  8.07 \\
Log scale accretion disk luminosity, in erg s$^{-1}$ & 45.39 \\
\hline
\end{tabular}
\end{table}

\section{Discussion and Conclusion}\label{sec:conclusion}
Using near-IR spectroscopy, we observed redshifted H$\beta$ emission line of TXS 1206+549 and obtained H$\beta$ FWHM of $1194\pm77$ km s$^{-1}$ and the flux ratio of [O III]5007 to H$\beta$ is low ($\sim 0.7$), suggesting TXS 1206+549 to be a NLS1. However, our spectrum of low signal-to-noise ratio (S/N) and limited wavelength coverage shows weak/no optical Fe II emission in TXS 1206+549. However, UV Fe II lines are seen in the SDSS optical spectrum. In terms of optical properties, it shows similarities with the $\gamma$-NLS1 PKS 2004$-$447 \citep{2001ApJ...558..578O} having weak Fe II emission rather than the other $\gamma$-NLS1 1H 0324+3410 \citep{2007ApJ...658L..13Z} having strong Fe II emission although both of them have similar H$\beta$ FWHM ($\sim$1600 km s$^{-1}$). However,  PKS 2004$-$447 exhibits a steep radio spectrum compared to the flat radio spectrum seen in TXS 1206+549 and 1H 0324+3410. We note that high S/N spectroscopic observations with a wider wavelength coverage will be helpful to quantify its optical Fe II strength and obtain detailed information about the shape of the H$\beta$ profile as the broad component of NLS1s are usually better represented by a Lorentzian profile rather than a Gaussian \citep[e.g.,][]{2001A&A...372..730V,2017ApJS..229...39R}. 

Using $\log L_{5100}=45.20$ erg s$^{-1}$ 
estimated from the power-law continuum and the H$\beta$ scaling relation 
of \citet{2006ApJ...641..689V}, we obtained 
$M_{\mathrm{BH}}=4.63\times 10^{7} \, M_{\odot}$ for TXS 1206+549. On the other hand, using Mg II line (Table \ref{Table:data}) we obtained  $M_{\mathrm{BH}}=2.97\times 10^{8} \, M_{\odot}$ \citep{2020ApJS..249...17R}.
However, black hole mass estimation based on continuum luminosity is likely to be uncertain, due to the non-thermal emission from the jet that contributes to the observed continuum flux 
\citep[e.g.][]{2020A&A...642A..59R}. The Mg II line equivalent width for TXS 1206+549 decreases by a factor of 3 between MJD = 52672 to 57430 (see Table \ref{Table:data}) as the Mg II flux remains almost unchanged while continuum luminosity at 3000\AA\ increases by 0.4 dex suggesting significant non-thermal contribution to the continuum that does not contribute in ionizing the line-emitting gas. Therefore, we also estimated the 
5100\AA\, continuum luminosity from the luminosity of the 
H$\beta$ using the scaling relation given by \citet{2020ApJS..249...17R} and obtained $\log L_{5100}=44.77$ erg s$^{-1}$ and a virial black hole mass of $M_{\mathrm{BH}}=2.8\times 10^{7} \, 
M_{\odot}$.  
Overall, the black hole mass of TXS 1206+549 is consistent with the  average black hole mass of NLS1s \citep[$<\log M_{\mathrm{BH}}>= 6.9\pm0.4 M_{\odot}$;][]{2017ApJS..229...39R} and within the range of black hole mass of $\gamma$-ray detected NLS1s ($<\log M_{\mathrm{BH}}>= 7.8\pm0.6 M_{\odot}$)
but lower than that of blazars \citep[$<\log M_{\mathrm{BH}}>= 8.9\pm0.5 M_{\odot}$;][]{2019ApJ...872..169P}. With a bolometric correction factor of 9.26 \citep{2006ApJS..166..470R}, the Eddington ratio is found to be 1.5 (using $\log L_{5100}=44.77$ erg s$^{-1}$), suggesting super Eddington accretion.

TXS 1206+549 is a radio-loud source showing a compact radio emission. The high amplitude of infrared variability in time scale of months to days suggests emission region is compact and significant non-thermal 
contribution from jet in the infrared. The SED shows a double hump structure typical for $\gamma$-ray emitting blazars (see Figure \ref{Fig:sed}). From our model fits (see Table \ref{tab:sed})
we found magnetic field strength and bulk Lorentz factor of 
3.2 Gauss and 11 respectively, similar to that found by \cite{2020_Tan}. The
magnetic field found for TXS 1206+549
is similar to that of other $\gamma$-ray 
emitting NLS1s \citep{2019ApJ...872..169P}.

NLS1s are usually considered to be low-luminous AGN ($\log L_{\mathrm{bol}}<46$ erg s$^{-1}$). However, with the discovery of more number of $\gamma$-ray detected NLS1s at high-$z$ and higher luminosity such as TXS 1206+549, they smoothly join the FSRQ branch of blazars. Identification of more such high-$z$ NLS1 and their broadband SED analysis will allow us to have a direct comparison of the physical properties of 
high-$z$ $\gamma$-ray detected NLS1s with that of high-$z$ blazars. Such
a study will allow one to probe the formation and evolution of radio-jets 
in NLS1s (sources with low black hole mass and high accretion 
rate relative to Eddington) {\it vis-a-vis} powerful blazars (sources
with high black hole mass and low accretion rate relative to Eddington).

\section*{Acknowledgements}     
We are thankful for the comments and suggestions by the anonymous referee, which helped us to improve the manuscript.  This
work is based on data collected at the Subaru Telescope, which is operated by the National Astronomical Observatory of Japan (NAOJ). Part of this work is based on archival data, software, or online services provided by the SPACE SCIENCE DATA CENTER (SSDC). This work was supported by JSPS KAKENHI Grant Number JP18H03717. JK acknowledges financial support from the Academy of Finland, grant 311438. JS was supported by Basic Science Research Program through the National Research Foundation of Korea (NRF) funded by the Ministry of Education (2019R1A6A3A01093189). SR thanks Neha Sharma for carefully reading the manuscript.

\section*{Data Availability}
The data underlying this article will be shared on reasonable request to the corresponding author.

\bibliographystyle{mnras}

\end{document}